\documentclass{ws-procs961x669}            

\usepackage{braket}

\newcommand{\be}{\begin{equation}}
\newcommand{\ee}{\end{equation}}
\newcommand{\bea}{\begin{eqnarray}}
\newcommand{\eea}{\end{eqnarray}}

\begin{document}
\title{{\bf Adiabatic regularization for massive and massless spin-$1$ fields}{\footnote{Expanded version of the talk given by F.J. M-G at the Seventeenth Marcel Grossmann Meeting (2024), to appear in the conference proceedings.}}}

\author{F. Javier Mara\~{n}\'on-Gonz\'alez$^*$ and Jos\'e Navarro-Salas$^\dagger$}

\address{Departamento de F\'isica Te\'orica and IFIC, Centro Mixto Universidad de Valencia-CSIC.\\
Facultad de F\'isica, Universidad de Valencia, Burjassot-46100, Valencia, Spain.\\
$^*$E-mail: jmarag@ific.uv.es
$^\dagger$E-mail: jnavarro@ific.uv.es}


\begin{abstract}

 The adiabatic regularization method is likely the most direct and intuitive renormalization 
scheme for
FLRW cosmologies. The method requires one to start with a nonvanishing mass, but massless theories can be studied by taking the massless limit at the end of the calculations. 
For spin-1, however, this limit changes the number of degrees of freedom, so it cannot be performed
directly. In this work, we show a direct approach that begins with the canonical quantization of the physical degrees of freedom of a massive
Proca field. We give the details of the construction and show that, in the massless limit, the
renormalized stress-energy tensor of the Proca field is closely related to that of a minimally coupled
scalar field. 
For completeness and pedagogical purposes, we also include a brief orientation to the adiabatic method for scalar fields. The construction of the adiabatic subtractions in momentum space is compared with that of the BPHZ  method for renormalizing Feynman loop integrals.

\end{abstract}


\bodymatter
\section{Introduction}

The phenomenon of particle creation by gravitational fields was originally discovered \cite{Parker} in the study of quantum fields in an expanding universe, typically described by a (spatially flat) Friedmann-Lema\^{i}tre-Robertson-Walker (FLRW) spacetime 
\be \label{FLRW}ds^2 = dt^2 - a^2(t) (dx^2 + dy^2 + dz^2) \ . \ee
A quantized scalar field  evolving in this background does not, in general, admit a unique preferred definition of the vacuum state. This is because the creation and annihilation operators of the quantized field, induced by the expansion factor $a(t)$, evolve into a combination of creation and annihilation operators. This fact leads to the fundamental prediction of particle production by the time-varying gravitational field. For recent reviews, see Refs. \cite{kolb-long, ford}.
In this context, one is immediately faced with the problem of  ultraviolet divergences that  appear in the evaluation of vacuum expectation values of quadratic field observables, such as the stress-energy tensor. The problem is more complex in a curved spacetime than in Minkowski space, due to the appearance of additional divergences induced by the curvature.
Because of the spatial translational symmetries of the background, the vacuum expectation values of the stress-energy tensor can be expressed as an integral in momentum space
\begin{equation}\label{Tformal}
\left\langle 0\left|T_{\mu \nu}\right| 0\right\rangle=\int d^3 k \ T_{\mu \nu}(\vec k, m;  a(t))\ , \end{equation}
where $m$ is the mass of the field. The above expression is formally divergent in the ultraviolet, and one should work out a method of subtracting the divergences to obtain a finite, physically consistent result. The idea of the renormalization method introduced in \cite{parker-fulling, parker-fulling2} (and called adiabatic regularization \footnote{For a historical perspective on the adiabatic method and particle creation, see \cite{parker12}. For textbook introductions, in the context of the general theory of quantum fields in curved space, see Refs. \cite{birrell-davies, fulling, parker-toms, hu-verdaguer}. The method was shown in \cite{parker-fulling} to be equivalent to the  n-wave regularization technique of Ref. \cite{ZS}.}) is to generate a subtraction algorithm directly for the integrand in (\ref{Tformal})
\begin{equation}\label{Tformal2}
\left\langle 0\left|T_{\mu \nu}\right| 0\right\rangle=\int d^3 k \ [T_{\mu \nu}(\vec k, m; a(t)) - (\text{Subtractions})_{\mu\nu} (\vec k)]
\ . \end{equation}
By construction, the renormalization method works in momentum space and does not depend on any regulator. The subtraction terms are systematically obtained by an adiabatic expansion of the quantities $T_{\mu \nu}(\vec k, a(t))$ with respect to the prescribed, external function $a(t)$. The subtraction procedure performed in this manner also functions as a ``regularization'' procedure. This justifies the traditional name of adiabatic regularization for the whole procedure. It is important to note that using the term ``regularization'' for a renormalization method that does not involve a regulator can be somewhat confusing. 
Usually, the regularization (e.g.  point-splitting, dimensional regularization, etc)  is disentangled  from the subtraction or renormalization. The regularized quantities are used to generate finite quantities on which the subtraction or renormalization can later be performed. In this sense, the adiabatic method is reminiscent of the BPHZ (Bogoliubov-Parasiuk-Hepp-Zimmermann) method for renormalizing Feynman integrals \cite{BPHZ, BPHZ2}. Neither method requires any prior regularization. 

The adiabatic method was further developed for scalar fields in Refs. \cite{Hu, Birrell78, bunch80, Anderson-Parker}, and has been used in a wide range of applications, such as in Refs. \cite{Anderson, Habit, agullo-navarro-olmo-parker, Bavarsad:2016cxh, Bavarsad:2017oyv, Moreno-Pulido20, Sola22, Silvia22, ferreiro-torrenti, silvia, Hu23}. A key point in the implementation of the adiabatic scheme to the scalar field is the WKB-type ansatz used to expand the mode functions before the adiabatic expansion of $T_{\mu\nu}(\vec k, m; a(t))$. The first terms of the later expansion fix the necessary subtractions. A conventional WKB approach is not possible for spin-$1/2$ fields, and an alternative strategy was needed. A self-consistent solution was found in Refs. \cite{LNT, RNT} (see also \cite{Ghosh15, Yukawa, BFNV}). The extension to fermionic fields has been further generalized to include external electromagnetic fields \cite{Ferreiro18, Ferreiro18b, BNP}, which can also be used in different applications \cite{Silvia-Paul, Eduardo-Luis}.

In general, the self-consistent implementation of the adiabatic method requires fields of non-zero mass\footnote{Note the parallelism with the BPHZ scheme. For massless fields the method was generalized in Ref. \cite{L76}, and is then called the BPHZL method. For the adiabatic regularization of massless scalar fields see \cite{Ferreiro-Navarroplb}.},  otherwise the method introduces  an artificial infrared divergence. 
In this context, a very important issue for the adiabatic method is how to obtain the conformal anomaly. The conformally invariant case can be obtained by performing the zero-mass limit 
 at the end of the calculations, as first given in Ref. \cite{Hu} for scalar fields and in Ref. \cite{LNT} for spin-$1/2$ fields. 
The extension of the adiabatic method for bosonic fields of spin-$1$ is expected to work well using WKB-type expansions. It was first explored in \cite{chimento-cossarini}. However the apparent inconsistency of the massless limit\footnote{A finite behavior for the adiabatic subtractions of the fourth order was (wrongly) expected, probably induced by comparison with the finite behavior of the fourth-order adiabatic subtractions for a conformally coupled scalar field. } was not understood, and the authors of Ref. \cite{chimento-cossarini} concluded that the adiabatic method required an additional ``gauge-breaking'' term to the Lagrangian of the Proca field\footnote{See also Refs. \cite{agullo, chu-koyama, ordines-carlson}.},  as well as a complex ghost field. One of the goals of this work is to clarify this issue and offer an improved and broader view. We  complement and expand the analysis given in Ref. \cite{Javier-Pepe23}.

The paper is organized as follows. In Sec. \ref{scalars} we give a general overview of the adiabatic method for scalar fields, with a general coupling $\xi$ to the scalar curvature. We find it useful to include an interesting  comparison between the adiabatic method and  the BPHZ renormalization method for dealing with divergent Feynman graphs. 
In Section \ref{spin1} we focus on  the spin-$1$  Proca field  and the corresponding WKB-type ansatz for the transverse and longitudinal polarizations. We give the details of  the construction of the adiabatic subtraction terms for the stress-energy tensor. We also discuss the massless limit. Our important observation is that  the divergences of the formal expression for $\langle T_{\mu\nu}^{\text{Proca}}\rangle$ coincide exactly with similar ultraviolet divergences found for a minimally coupled scalar field (when the mass is set to zero at the end of the calculation). As a consequence, taking the difference  between  the renormalized quantities 
 $\langle T^{\text{Proca}}_{\mu\nu} \rangle_{\text{ren}}$ and  $\langle T^{\text{Scalar }\xi=0}_{\mu\nu}\rangle_{\text{ren}}$, 
and setting at the end $m=0$, one exactly gets the renormalized stress-energy tensor of the Maxwell field $\langle T^{\text{Maxwell}}_{\mu\nu} \rangle_{\text{ren}}$.  We consider this to be  a very significant  property of the adiabatic regularization description of a Proca field. It  fills an important  gap in the existing literature on the adiabatic method. In Section \ref{discussion} we summarize our work and discuss some of the implications of our main result.

 Throughout this work, we use units in which $\hbar = c = 1$. Our  conventions for the signature of the metric and the curvature tensor follow Refs. \cite{birrell-davies, parker-toms}.

\section{Brief orientation to the adiabatic method for spin-$0$ fields. Comparison with the BPHZ renormalization method} \label{scalars}

 Let us consider a scalar field $\phi$ propagating in  a spatially flat FLRW background (\ref{FLRW}). The  action functional is given by
 \be S= \int \sqrt{-g}d^4x \, \frac{1}{2}( g^{\mu\nu} \partial_\mu \phi \partial_\nu \phi -m^2 \phi^2 -\xi R \phi^2) \ . \ee
 The field obeys the  field equation
\be (\Box + m^2 + \xi R) \phi =0 \ . \ee
To implement the canonical quantization in the assumed FLRW spacetime, we expand the field  into a complete  set of orthonormal solutions
\be \label{modedec}\phi (t,\vec{x}) = \int \mathrm{d}^3 k \, \left(a_{\vec k} f_{\vec k} (t,\vec{x}) + a_{\vec k}^{\dagger}f^*_{\vec k} (t,\vec{x})\right) \ . \ee
It is convenient to factorize the dependence on the spatial coordinates $\vec x$ and time as follows
\be f_{\vec k}(t,\vec{x}) = \frac{1}{\sqrt{2(2\pi)^3 a^3(t)}} h_k(t) e^{i\vec k\cdot\vec x} \ , \ee
where $h_k(t)$ obeys the equation (here $k\equiv |\vec k|$)
\be \label{waveq}\ddot h_k + (\omega_k^2  + \sigma) h_k =0 \ . \ee
The physical frequency $\omega_k$ in the expanding universe is defined as $\omega_k = \sqrt{k^2/a^2 + m^2}$, while $\sigma$ is given by 
\be \sigma = (6\xi -3/4) \dot a^2/a^2 +(6\xi -3/2)\ddot a/a \ . \ee  The field modes have been assumed to be orthonormal with respect to the Klein-Gordon inner product
\be (\phi_1, \phi_2) = i\int d^3 x a^3 (\phi_1^* \partial_t \phi_2 - \partial_t \phi_1^* \phi_2) , \ee
implying that  the functions $h_k$ must satisfy the normalization condition
\be \label{wronskian} h_k \dot h_k^* - h_k^* \dot h_k = 2i \ . \ee

To illustrate how the adiabatic subtraction method works, let us consider the formal expression for the two-point function. Since the vacuum is defined by the usual condition $a_{\vec k} |0\rangle =0$, one easily obtains (from now on we omit writing explicitly the vacuum state in vacuum expectation values)
\be \label{phi2} \langle \phi^2(t,\vec{x})\rangle = \frac{1}{16\pi^3 a^3(t)}\int \mathrm{d}^3k \,  |h_k(t)|^2 \ . \ee
This is an integral in momentum space, where the integrand function $h_k(t)$ depends on $k, m$.
$h_k(t)$ also has an implicit  non-local dependence on the expansion factor $a(t)$.

\subsection{BPHZ subtractions}
To better understand the construction of the adiabatic subtractions we find useful to sketch here the construction of the subtraction terms in the BPHZ method of renormalization\footnote{The method is equivalent to the usual method based on counterterms.}. For simplicity, let us  consider a  typical one-loop integral in the $\phi^4$ theory (describing the scattering $p_1 + p_2 \to p_3 + p_4$ in the t-channel) \cite{Peskin}
 \be  \int \frac{d^4k}{(2\pi)^4} \frac{1}{k^2 -m^2 +i\epsilon} \frac{1}{(k+p_1-p_3)^2-m^2 + i\epsilon} 
 \ . \ee
  In Euclidean space the above integral is given by 
\be \label{J2}
J(p\equiv p_1-p_3) =\int \frac{d^4 k}{(2\pi)^4} \frac{1}{\left[k^2+m^2\right]\left[(p+k)^2+m^2\right]} \equiv  \int d^4k \ I(p, k) \ .
\ee
This integral is logarithmically divergent, with a superficially  degree of divergence $D=0$. A finite part for such an integral  can be extracted in different manners. 

In general, for an integral $J(p) \equiv \int d^4 k I(p, k)$ associated to a loop graph depending on the external momentum $p$, and superficially degree of divergence $D=N$, the BPHZ momentum space subtraction is defined by (see, for example, Ref. \cite{piguet-sorella})
\bea 
 J^{\text{BPHZ}}(p) = \int \ d^4k \ [I(p, k)&-& [I(0, k)+p^\mu \frac{\partial I}{\partial p^\mu}(0, k)+\cdots \nonumber  \\ &+&\frac{1}{N !} p^{\mu_1}\cdots p^{\mu_N} \frac{\partial^N I}{\partial p^{\mu_1} \cdots \partial p^{\mu_N}}(0, k) ]]  \ ,  \eea 
 which represents a convergent integral. 
 Extracting a finite part of $J(p)$ by means of this BPHZ prescription yields the expression $J^{\text {BPHZ}}(p)$, which is determined only up to a polynomial in $p$ of degree $D=N$.
 In configuration space this corresponds to finite local counterterms involving derivatives of the field.
 
Returning to our case, for the integral $J(p)$ in (\ref{J2}), according to the BPHZ rule, and since $D=0$, we get a convergent integral
\be 
\begin{aligned}
J^{\mathrm{BPHZ}}(p) =\int \frac{d^4 k}{(2\pi)^4} \left(\frac{1}{\left[k^2+m^2\right]\left[(p+k)^2+m^2\right]}-\frac{1}{\left[k^2+m^2\right]^2}\right)
\end{aligned}
\ . \ee
This example shows that subtractions are applied to the integrand. No regularization is introduced. We could have used a regularization procedure to evaluate each integral separately (e.g., a cut-off, dimensional regularization, etc). Then subtract the finite integrals and remove the regularization parameter at the end. The final result is the same.

\subsection{Adiabatic subtractions}

With the BPHZ procedure in mind, it is easier to understand the construction of the adiabatic subtractions to convert (\ref{phi2}) into a finite integral. Instead of the external momenta $p$  in the Feynman loop integrals, in our problem  we have, as the external agent, the cosmic factor $a(t)$.  Therefore, the Taylor expansion around $p^\mu \sim 0$ should naturally be replaced by an expansion around $\dot a (t) \sim 0$, $\ddot a(t) \sim 0$, $\dddot a (t) \sim 0$, and so on.  This corresponds to the  concept of  {\it adiabatic expansion} for $a(t)$. The adiabatic order in each term of the expansion is determined by the number of time derivatives in $a(t)$. For example, $\dot{a}$ is of adiabatic order 1, $\ddot a$ and $\dot a^2$ are of adiabatic order 2, and $\dot{a} \dot{a}^2$ is of adiabatic order 3.   

To evaluate the adiabatic expansion of $|h_k(t)|^2$, i.e., the integrand in (\ref{phi2}), we  first construct the adiabatic expansion of the mode function $h_k(t)$. To this end it is useful to consider the WKB-type ansatz\footnote{The assumption of this ansatz for the modes defines a family of states referred as  adiabatic vacua. This condition is equivalent, in a FLRW spacetime, to the so-called Hadamard condition \cite{pirk, sergi}.} 

\be \label{WKB} h_k(t) \sim \frac{1}{\sqrt{W_k(t)}} e^{-i\int^t W_k(t') \mathrm{d}t'} \ , \ee
where the leading term for $W_k(t)$ is the physical frequency $\omega_k$. Hence for  $\dot a(t)=0$ the modes behave as the Minkowskian modes. The higher-order terms (we use the notation $W_k^{(n)}\equiv \omega^{(n)}, n \geq 1$, compared to Ref. \cite{parker-toms}, section 3.1)
\be \label{W} W_k(t) \sim \omega_k + {\color{black} W_k^{(1)}} + {\color{black} W_k^{(2)}} + \cdots  \ee
are systematically obtained 
by  plugging the ansatz (\ref{WKB}) into the field equation for $h_k(t)$ obtained in (\ref{waveq}).  The leading term  determines univocally the higher-order terms in the adiabatic expansion. One gets 
\be W_k^{(1)} \label{Wadiabatic} =0 \ , \ee
\be 
W_k^{(2)} = \frac{\sigma}{2 \omega}-\frac{\ddot \omega}{4 \omega^2}+\frac{3 \dot \omega^2}{8 \omega ^3}
\ , \ee
\be
W_k^{(3)}=0
\ , \ee
\be
\begin{split}
W_k^{(4)} = &-\frac{\ddot \sigma }{8 \omega  ^3}+\frac{5 \dot \sigma 
   \dot \omega}{8 \omega  ^4}+\frac{3 \sigma  \ddot \omega }{8 \omega  ^4}-\frac{19 \sigma  \dot \omega
   ^2}{16 \omega  ^5}-\frac{\sigma  ^2}{8 \omega
    ^3}+\frac{\ddddot \omega }{16 \omega
    ^4} \\ 
    &-\frac{13 \ddot \omega ^2}{32 \omega
    ^5}-\frac{297 \dot\omega  ^4}{128 \omega
    ^7}-\frac{5 \dddot\omega  \dot\omega }{8 \omega
    ^5}+\frac{99 \dot\omega ^2 \ddot \omega}{32 \omega
    ^6}
\ , \end{split}
\ee
where we have used the simplifying notation $\omega_k\equiv \omega$.  We do not write the explicit form of the time derivatives of $\omega$ and $\sigma$. For example, $\dot \omega = -k^2\dot a/a^3\omega$, etc.   We note that all the terms of odd adiabatic order are zero. 

We can continue the procedure by computing the adiabatic expansion of the composed function $|h_k(t)|^2$. We get 
\be \label{expansionphi2} |h_k(t)|^2 \sim W_k^{-1} = \omega^{-1} + (W_k^{-1})^{(1)}+ (W_k^{-1})^{(2)} + (W_k^{-1})^{(3)} +\cdots  \ .  \ee
 Similarly to the case of the BPHZ algorithm, the above adiabatic expansion  captures the ultraviolet  divergences in the $k$-integral for the two-point function in (\ref{phi2}). These divergences are independent of the  specific vacuum state, defined by a specific choice of the mode functions $h_k(t)$.

 The required renormalization subtractions are then obtained from (\ref{expansionphi2}), in full analogy with the BPHZ method.  The subtraction terms are given by  the first terms in (\ref{expansionphi2}). The role of the superficial degree of divergence $D$ of the Feynman loop integral in the BPHZ method is replaced here by the adiabatic order of the corresponding operator. The operator $\phi^2$  has  adiabatic order $2$. [The terms of third adiabatic order and beyond   are all finite.] Therefore, the rule is to subtract the full  adiabatic terms of order $0$, $1$ and $2$.  The renormalized expression for $\langle \phi^2\rangle_{\rm ren}$ is then constructed by the momentum integral
\be \label{phi2ren} \langle \phi^2\rangle_{\rm ren} = \frac{1}{4\pi^2 a^3}\int_0^\infty \!\!\!\! \mathrm{d}k\, k^2 \left( |h_k|^2 -   \omega^{-1} - (W_k^{-1})^{(1)}- (W_k^{-1})^{(2)}  \right) \ . \ee
The final result is (some terms can be integrated exactly)
\be 
\left\langle\phi^2\right\rangle_{\text {ren }}=\frac{1}{4 \pi^2 a^3} \int_0^{\infty} \!\!\!\! \mathrm{d}k\, k^2\left[\left|h_k\right|^2-\frac{1}{\omega}-\frac{\left(\frac{1}{6}-\xi\right) R}{2 \omega^3}\right]-\frac{R}{288 \pi^2}
\ , \ee
where we have used that $R=6(\dot a^2/a^2 + \ddot a/a)$. Note that the operator $\phi^2$ is of adiabatic order $2$, although for $\xi=1/6$ all terms of the integrand of second adiabatic order (i.e., proportional to $\dot a^2$ or $\ddot a$) are UV finite. Again, the parallelism with the BPHZ rules is maintained since, in that language, it is the superficial degree of divergence (and not the proper degree of divergence) what determines the subtraction terms.

The stress-energy operator $T^{\mu\nu}$ is of  
adiabatic order $4$. Therefore, 
the renormalization of $\langle T^{\mu\nu} \rangle$ requires subtracting until and including the fourth adiabatic order. An important issue here is the conformal anomaly for massless fields. The adiabatic subtractions need massive fields. To illustrate the point, let us consider the trace of the stress-energy tensor. The operator expression of the trace, once the equations of motion have been used, can be written as follows
\be \label{truetrace}T^\alpha_\alpha = (6 \xi - 1) \partial^\alpha \phi \partial_\alpha \phi + \xi (1- 6 \xi)R \phi^2 + (2-6 \xi) m^2 \phi^2 \ .  \ee
For conformal coupling $\xi=1/6$, and for massless fields the trace is zero. This shows the conformal invariance of the classical theory for these values of the parameters. For the adiabatic renormalization we need $m\neq 0$. The massless limit must be taken at the end of the calculation.
For $\xi=1/6$ the vacuum expectation values of the trace reduces to
\be \langle T^\alpha_\alpha \rangle = m^2 \langle\phi^2 \rangle \ . \ee
The adiabatic subtraction terms to fourth order are finite in the limit $m^2\to 0$ and they define the conventional trace anomaly of a massless and conformally coupled scalar field 

\bea
\langle T^\alpha_\alpha \rangle_{\text{ren}} &=& \lim_{m\to 0} -m^2 \langle\phi^2 \rangle^{(4)} \nonumber \\
&=&\left(4 \pi^2 a^3\right)^{-1} \lim _{m \rightarrow 0} \int_{\dot{\phi}}^{\infty} d k k^2 \left[\frac{7 m^4 \ddot{a}^2}{16 a^2 \omega^7}+\frac{m^4 \ddddot a}{16 a \omega^7}
+\frac{m^4 \dot{a}^4}{2 a^4 \omega^7} \right.  \nonumber \\ &+&\frac{11 m^4 \dot{a} \dddot{a}}{16 a^2 \omega^7}+\frac{33 m^4 \ddot a^2 \ddot{a}}{16 a^3 \omega^7}-\frac{21 m^6 \ddot a^2}{32 a^2 \omega^9}-\frac{49 m^6 \dot a^4}{8 a^4 \omega^9} 
-\frac{7 m^6 \dot{a} \dddot a}{8 a^2 \omega^9} \nonumber \\&-&\left.\frac{35 m^6 \dot{a}^2 \ddot{a}}{4 a^3 \omega^9}+\frac{231 m^8 \dot{a}^4}{16 a^4 \omega^{11}}+\frac{231 m^8 \dot{a}^2 \ddot{a}}{32 a^3 \omega^{11}}-\frac{1155 m^{10} \dot{a}^4}{128 a^4 \omega^3}\right]
\eea
All  integrals are finite, as is the limit. One obtains a covariant result \cite{parker-toms}
\begin{equation}
\begin{split}
    \langle T^\alpha_\alpha \rangle_{\text{ren}} 
&= \frac{1}{2880\pi^2} \left[-\left(R_{\mu\nu}R^{\mu\nu} -\frac{1}{3}R^2\right) + \Box R\right]\ .
\end{split}
\end{equation}

\section{Adiabatic method for spin-$1$ fields.}\label{spin1}

The action for the Proca {\color{black}fields} in curved space is
\be \label{Sproca} S= \int d^4x \sqrt{-g}\left(-\frac{1}{4}F_{\mu\nu}F^{\mu\nu} +\frac{1}{2}m^2 A^\mu A_\mu\right) \ , \ee
implying the  field equations
\be \label{feqsP}\nabla_\mu F^{\mu\nu} + m^2 A^\nu =0 \ . \ee
The above  equations enforce, for $m^2 \neq 0$, the scalar constraint
\be \label{subc}\nabla_\nu A^\nu=0 \ . \ee 
In addition, the inner product for the Proca field in curved spacetime takes the form \cite{parker-wang} 
\bea \label{ipPcs} (A_1, A_2) &=&  -i \int_\Sigma d\Sigma^\mu g_{\rho\nu}[A_{1}^{*\rho} \nabla_\mu A_{2 }^\nu - (\nabla_\mu A_{1}^{*\rho})  A_{2 }^\nu] \ , \eea
where $\Sigma$ is the initial-time Cauchy hypersurface and  $d\Sigma^\mu = n^\mu d\Sigma$. $d\Sigma$ is the volume element on $\Sigma$ and $n^\mu$ is a future-directed unit vector orthogonal to  $\Sigma$.

In the FLRW spacetime $ds^2 = dt^2 - a^2(t) (dx^2 + dy^2 + dz^2)$ we have 
$ \nabla_0A^0 = \partial_0 A^0$, $\nabla_0 A^j =  \partial_0 A^j +\frac{\dot a}{a}A^j$. Therefore, 
the inner product can be expressed as 
\bea \label{ipPcs2} (A_1, A_2) &=&  
 -i \int d^3x a^3 \{[A_{1}^{*0} \partial_0 A_{2 }^0 - (\partial_0 A_{1}^{*0})  A_{2 }^0] \nonumber \\
 &-& \sum_ja^2 [A_{1}^{*j} \partial_0 A_{2 }^j - (\partial_0 A_{1}^{*j})  A_{2 }^j] \}\ . \eea

 \subsection{Modes for the Proca field}
The explicit form of the inner product suggests a natural  ansatz for the modes of the Proca field  as
\be \label{Procamodeansatz} A^{\mu (r)}_{\vec k} =    \frac{e^{i\vec k \vec x}}{a\sqrt{2(2\pi a)^3 }} \epsilon ^{\mu (r)}(\vec k, t)  f_k^{(r)}(t).  \ee 
Where $r=1,2$ denote the transverse modes and $r=3$ the longitudinal ones. We will also rename $f_k^{(3)} = l_k$ as the longitudinal mode functions, and similarly for the transverse $f_k^{(1)} = f_k^{(2)} = h_k$. In \cite{Javier-Pepe23} was proved that a basis of vectors $\epsilon^{\mu(r)}$ can be chosen so that the above modes are properly normalized with respect to the inner product
\be (A^{(r)}_{\vec k}, A^{(s)}_{\vec k'}) = \delta^{3}(\vec k -\vec k')\delta^{rs} \ , \ee
and the field equations become (\ref{waveq}) with different choices of $\sigma(t)$ for the two sets of polarizations
\begin{equation} \label{sigmas}
\begin{split}
    \sigma_h &= \left(\frac{1}{4}\right) \frac{\dot a^2}{a^2} + \left(-\frac{1}{2}\right) \frac{\ddot a}{a} \\
    \sigma_l &= \left(-\frac34 + \frac{4 m^2}{ \omega_k^2} - \frac{3 m^4}{\omega_k^4}\right) \left(\frac{\dot a}{a}\right)^2
+ \left( - \frac32 + \frac{m^2}{\omega_k^2}\right) \frac{\ddot a}{a} \ .
\end{split}
\end{equation}
These reproduce, in the massless limit, the equations of motion of a conformal scalar field for the transverse polarizations, and a minimally coupled scalar field for the longitudinal one.

An explicit realisation of a suitable basis of vectors is
\be \epsilon^{\mu (1)}(\vec k)= \frac{1}{\sqrt{k_1^2 + k_2^2}}\left(0, -k_2, k_1, 0\right)\ , \ee
\be \epsilon^{\mu (2)}(\vec k)= \frac{k_3}{k \sqrt{k_1^2+k_2^2}}\left(0, -k_1, -k_2, \frac{k_1^2 + k_2^2}{k_3} \right)\ , \ee
\be \epsilon^{\mu (3)}(\vec k, t)= \frac{\omega_k}{k m}\left(\frac{k^2 \mathcal{W}_k}{\omega_k^2}, \vec k\right) \ , \ \ \ \mathcal{W}_k = i \left( \frac{\dot l_k}{l_k} - \frac{3\dot a}{2a} +\frac{m^2 \dot a}{\omega_k^2 a} \right)  \ , \ee
where $\vec k = (k_1, k_2, k_3)$. One can then perform a mode expansion of the field operator
\be A^\mu = \sum_{r=1}^3\int d^3k \left(A_{\vec k}^{(r)} \, \epsilon^{\mu {(r)}} \, a^r_{\vec k}  + \text{h.c.} \right) \ , \ee
where
\be [a^r_{\vec k}, a^{s\dagger}_{\vec k'}]= \delta^{rs}\delta^3(\vec k -\vec k') \ .\ee

\subsection{Adiabatic expansion of field modes and  stress-tensor components}
The ansatz (\ref{Procamodeansatz}) with the specified vector basis is compatible with the WKB expansions
\be  h_k(t) \sim \frac{1}{\sqrt{\Omega_k(t)}} e^{-i\int^t \Omega_k(t') \mathrm{d}t'} \ , \ee
\be l_k(t) \sim \frac{1}{\sqrt{\Lambda_k(t)}} e^{-i\int^t \Lambda_k(t') \mathrm{d}t'} \ . \ee

As a result, once the relevant physical quantities have been expressed in terms of $\Omega_k$ and $\Lambda_k$, an adiabatic expansion can be performed based on (\ref{Wadiabatic}) and (\ref{sigmas}). For example, the operator expression for the stress energy tensor is
\begin{equation}
    T_{\alpha\beta} = -F_\alpha^{\,\, \mu} F_{\beta \mu} + \frac{1}{4} g_{\alpha\beta} F_{\mu\nu} F^{\mu\nu} - m^2 \left( \frac{1}{2} g_{\alpha\beta} A^\mu A_\mu + A_\alpha A_\beta \right) \ .
\end{equation}
One can evaluate the above operator in the  vacuum as $\langle T_{\alpha\beta}\rangle$. The off-diagonal terms are odd in momentum-space, so they do not contribute. The space-space components can be shown to be equivalent by a suitable change of dummy integration (momentum) variables, hence resulting in the required isotropic pressures. The contributions from transverse ($\Omega$) and longitudinal ($\Lambda$) modes can be separated. In terms of the mode functions, one has 
\begin{equation}\label{rhoomega}
    \braket{\rho}_\Omega = \braket{T^0_0}_\Omega = \int \frac{d^3 k}{16 \pi^3 a^3}\left( -\frac{h^* \dot a \dot h}{2
   a}-\frac{h \dot a
  \dot h^*}{2
   a}+\frac{h h^*
   \dot a^2}{4 a^2}+\dot h
   \dot h^*+h h^* \omega
   ^2 \right)
\end{equation}
\begin{equation}\label{Pomega}
    \braket{P}_\Omega = -\braket{T^1_1}_\Omega = \int \frac{d^3 k}{48 \pi^3 a^3}\left(-\frac{h^* \dot a \dot h}{2
   a}-\frac{h \dot a
   \dot h^*}{2
   a}+\frac{h h^*
   \dot a^2}{4 a^2}+\dot h \dot
   h^* - 2 m^2 h
   h^* + h h^* \omega
   ^2 \right)
\end{equation}
\begin{equation}\label{rholambda}
\begin{split}
    \braket{\rho}_\Lambda = \braket{T^0_0}_\Lambda = \int \frac{d^3 k}{32 \pi^3 a^3}&\left( \frac{m^2 l^* \dot a
   \dot l}{a \omega ^2}-\frac{3
   l^* \dot a \dot l}{2
   a}+\frac{m^2 l
   \dot a \dot l^*}{a \omega
   ^2}-\frac{3 l \dot a
   \dot l^*}{2 a}+\frac{m^4
   l l^*
   \dot a^2}{a^2 \omega
   ^4}\right. \\
   &\left.-\frac{3 m^2 l
   l^* \dot a^2}{a^2
   \omega ^2}+\frac{9 l l^*
   \dot a^2}{4 a^2}+\dot l
   \dot l^*+l l^* \omega
   ^2 \right)
\end{split}
\end{equation}
\begin{equation}\label{Plambda}
\begin{split}
    \braket{P}_\Lambda = -\braket{T^1_1}_\Lambda = \int \frac{d^3 k}{96 \pi^3 a^3}&\left( -\frac{2 m^4 l^* \dot a
   \dot l}{a \omega ^4}+\frac{6
   m^2 l^* \dot a
   \dot l}{a \omega ^2}-\frac{9
   l^* \dot a \dot l}{2
   a}-\frac{2 m^4 l
   \dot a \dot l^*}{a \omega
   ^4} \right. \\
   &\left.+\frac{6 m^2 l
   \dot a \dot l^*}{a \omega
   ^2}-\frac{9 l \dot a
   \dot l^*}{2 a}-\frac{2 m^6
   l l^*
   \dot a^2}{a^2 \omega
   ^6}+\frac{9 m^4 l
   l^* \dot a^2}{a^2
   \omega ^4}\right. \\
   &\left.-\frac{27 m^2 l
   l^* \dot a^2}{2 a^2
   \omega ^2}+\frac{27 l l^*
   \dot a^2}{4 a^2}-\frac{2 m^2
   \dot l \dot l^*}{\omega ^2}+3
   \dot l \dot l^*-l
   l^* \omega ^2 \right) \ .
\end{split}
\end{equation}
A  detailed expression using the WKB ansatz is given in Appendix \ref{apA}.
The above expressions, expanded  up to fourth adiabatic order,  are given below. Since they contain all the divergences, they are all the required subtraction terms. [Note that finite integrals have been evaluated explicitly.]
\begin{equation}
\begin{split}
    \braket{\rho}_\Omega^{(0-4)} &= \frac{\ddot a^2}{480 \pi ^2
   a^2}+\frac{m^2 \dot a^2}{48 \pi ^2
   a^2}+\frac{\dot a^4}{240 \pi ^2
   a^4}-\frac{\dddot a
   \dot a}{240 \pi ^2
   a^2}-\frac{\dot a^2
   \ddot a}{240 \pi ^2 a^3} \\ 
   &+\int \frac{d^3 k}{8 \pi^3 a^3} \cdot \omega
\end{split}
\end{equation}
\begin{equation}
\begin{split}
    \braket{P}_\Omega^{(0-4)} &= \frac{\ddddot a}{720 \pi ^2
   a}-\frac{m^2 \ddot a}{72 \pi ^2
   a}+\frac{\ddot a^2}{480 \pi ^2
   a^2}-\frac{m^2 \dot a^2}{144 \pi ^2
   a^2}+\frac{\dot a^4}{720 \pi ^2
   a^4}+\frac{\dddot a
   \dot a}{360 \pi ^2
   a^2}\\
   &-\frac{\dot a^2
   \ddot a}{180 \pi ^2 a^3} 
   +\int \frac{d^3 k}{24 \pi^3 a^3} \left( \omega -\frac{m^2}{\omega } \right)
\end{split}
\end{equation}
\begin{equation}
\begin{split}
    \braket{\rho}_\Lambda^{(0-4)} &= -\frac{3 \ddot a^2}{320 \pi ^2
   a^2}+\frac{m^2 \dot a^2}{96 \pi ^2
   a^2}-\frac{7 \dot a^4}{240 \pi ^2
   a^4}+\frac{3 \dddot a
   \dot a}{160 \pi ^2 a^2}+\frac{13
   \dot a^2 \ddot a}{160 \pi ^2
   a^3} \\ 
   &+\int \frac{d^3 k}{16 \pi^3 a^3} \left( \frac{\ddot a^2}{8 a^2 \omega
   ^3}-\frac{m^2 \dot a^2}{2
   a^2 \omega ^3}+\frac{3
   \dot a^4}{8 a^4 \omega
   ^3}+\frac{\dot a^2}{2 a^2
   \omega }-\frac{\dddot a
   \dot a}{4 a^2 \omega
   ^3}-\frac{\dot a^2 \ddot a}{4
   a^3 \omega ^3}+\omega  \right) 
\end{split}
\end{equation}
\begin{equation}
\begin{split}
    \braket{P}_\Lambda^{(0-4)} &= -\frac{\ddddot a}{160 \pi ^2
   a}-\frac{m^2 \ddot a}{144 \pi ^2
   a}-\frac{13 \ddot a^2}{320 \pi ^2
   a^2}-\frac{13 m^2 \dot a^2}{288 \pi
   ^2 a^2}+\frac{31 \dot a^4}{1440 \pi ^2
   a^4}-\frac{13 \dddot a
   \dot a}{240 \pi ^2 a^2}\\
   &+\frac{13
   \dot a^2 \ddot a}{720 \pi ^2
   a^3}
   +\int \frac{d^3 k}{48 \pi^3 a^3} \left( \frac{\ddddot a}{4 a \omega
   ^3}+\frac{m^2
   \ddot a}{a \omega
   ^3}+\frac{3 \ddot a^2}{8 a^2
   \omega ^3}-\frac{\ddot a}{a
   \omega }+\frac{m^2
   \dot a^2}{a^2 \omega
   ^3}\right.\\
   &\left.+\frac{3 \dot a^4}{8 a^4
   \omega ^3}+\frac{\dot a^2}{2
   a^2 \omega
   }+\frac{\dddot a \dot a}{2
   a^2 \omega ^3}-\frac{3
   \dot a^2 \ddot a}{2 a^3
   \omega ^3}-\frac{m^2}{\omega }+\omega \right) \ .
\end{split}
\end{equation}
As can be seen, the longitudinal polarizations are more divergent. They diverge up to fourth adiabatic order, similarly to a minimally coupled scalar field\footnote{A detailed analysis of the adiabatic method applied to $\xi$-generic scalar fields can be found, in conformal time, in \cite{bunch80}. In appendix B we give the relevant expressions  in cosmic time.}. For a more detailed discussion of this relation, see \cite{Javier-Pepe23}. 
Since the stress-energy tensor is an operator of  adiabatic order $4$, we should always subtract until the fourth adiabatic order. Note again the similarity with the BPHZ rules for selecting  the subtraction terms.

\subsection{Relation to Maxwell theory}

The theory of classical electrodynamics (i.e. classical Maxwell theory) in a curved spacetime can be shown to be conformally invariant. After quantization, the theory develops a trace anomaly as the sole contribution to $\braket{T_{\mu\nu}}_\text{ren}g^{\mu\nu}$. This anomaly has been derived with many different approaches (see Refs. \cite{birrell-davies,parker-toms,Duff}) and is a well-established  result of Quantum Field Theory in Curved Spacetimes. 

For quantum theories with classical conformal invariance, it is possible to derive the whole renormalized stress energy tensor just from the trace anomaly in FLRW spacetimes. The procedure is based in the existence of a conformal Killing vector field in FLRW [see Ref. \cite{parker-toms} for a detailed explanation]. If the renormalized trace of a theory can be expressed as
\begin{equation}
    \braket{T_{\mu\nu}}_\text{ren}g^{\mu\nu} = \frac{1}{2880 \pi^2} \left[ \alpha \left(R_{\mu\nu}R^{\mu\nu}- \frac{1}{3}R^2\right) + \beta \, \Box R \right] \ ,
\end{equation}
then the most generic renormalized stress-energy tensor is
\begin{equation}
    \braket{\rho}_\text{ren} = -\frac{\beta  \ddot a^2}{960 \pi ^2
   a^2}-\frac{\alpha  \dot a^4}{960 \pi ^2
   a^4}-\frac{\beta  \dot a^4}{320 \pi ^2
   a^4}+\frac{\beta  \dddot a
   \dot a}{480 \pi ^2 a^2}+\frac{\beta 
   \dot a^2 \ddot a}{480 \pi ^2
   a^3}+\frac{E}{a^4}
\end{equation}
\begin{equation}
\begin{split}
    \braket{P}_\text{ren} = &-\frac{\beta  \ddddot a}{1440 \pi ^2
   a}-\frac{\beta  \ddot a^2}{960 \pi ^2
   a^2}-\frac{\alpha  \dot a^4}{2880 \pi ^2
   a^4}-\frac{\beta  \dot a^4}{960 \pi ^2
   a^4}-\frac{\beta  \dddot a
   \dot a}{720 \pi ^2 a^2}+\frac{\alpha 
   \dot a^2 \ddot a}{720 \pi ^2
   a^3}\\
   &+\frac{\beta  \dot a^2
   \ddot a}{240 \pi ^2 a^3}+\frac{E}{3
   a^4} \ ,
\end{split}
\end{equation}
with $E$ being an integration constant that can be interpreted as a classical energy-density. For an adiabatic vacuum $E=0$. The above relations are valid for conformal scalar fields ($\alpha = -1$, $\beta=1$) and Maxwell theory ($\alpha=-62$, $\beta=-18$). We note that the coefficient $\beta$ is intrinsically ambiguous. The adiabatic method agrees with the result $\beta=-18$, obtained by point-splitting and zeta-function regularization. 

Returning to the adiabatic subtractions for the Proca field derived in this paper, we can construct the renormalized stress energy tensor  as
\begin{equation}
    \braket{T_{\mu \nu}}_\text{ren} \equiv \braket{T_{\mu \nu}}_\text{modes} - \braket{T_{\mu \nu}}^{(0-4)} \ .
\end{equation}
If we further assume that there is no finite contribution from $\braket{T_{\mu \nu}}_\text{modes}$, then we can use our expressions  to obtain
\begin{equation}\label{main}
    \lim_{m\to0} \left(\braket{T_{\mu\nu}}_\text{ren}^\text{Proca} - \braket{T_{\mu\nu}}_\text{ren}^{\xi=0\text{ scalar}} \right) = \braket{T_{\mu\nu}}_\text{ren}^\text{Maxwell} \ ,
\end{equation}
since the divergent terms in the adiabatic subtractions will exactly cancel those of the modes in the massless limit. 

It is worth noting that, as an intermediate result, the transverse polarizations behave like two conformal scalar fields
\begin{equation}
    \lim_{m\to0} \braket{T_{\mu\nu}}_\text{ren}^{\text{Proca, }\Omega}= 2 \, \braket{T_{\mu\nu}}_\text{ren}^{\xi=1/6\text{ scalar}}  \ .
\end{equation}
Note that this with (\ref{main}) implies that, in the ($m\to 0$) quantum theory, neither the two transverse polarizations behave as quantum electromagnetism nor the longitudinal polarization behaves as a minimal scalar field.

\section{Summary and conclusions}\label{discussion}

In order to make the methods used in this paper accessible to a broad readership, we have included a brief introduction to the adiabatic method for scalar fields. To provide an intuition of how the renormalization subtractions work, we have compared the adiabatic technique with the BPHZ subtraction method. The subtractions of both methods are applied directly to the integrands of the divergent momentum space integrals. Thus, despite the traditional term ``adiabatic regularization'', there is no need to introduce any prior regularization.

In this work we have described the adiabatic method for renormalizing the stress-energy tensor of a Proca field in a FLRW spacetime. We have used the canonical quantization route to quantize the theory and evaluate the required adiabatic subtractions. Therefore, we avoid dealing with unphysical degrees of freedom and ghost fields. The method clearly demonstrates how the renormalized stress-energy tensor in the massless limit relates to that of the Maxwell field.
Furthermore, as explained in Ref. \cite{Javier-Pepe23}, the main result  (\ref{main}) is fully consistent with the approach based on the effective action \cite{buchbinder-shapiro21, buchbinder-berredo-shapiro} and fits  well with  established  results in de Sitter space \cite{allen-folacci}.

At the level of the classical field equations, one can say that the Proca field is equivalent, in the massless limit of (\ref{sigmas}), to two conformally invariant scalars (transverse polarizations) and one minimally coupled scalar (longitudinal polarization). 
This is not quite correct in the quantized theory, since the  renormalization is more sensitive to 
the true field content and distinguishes between the two descriptions.
In the same way, the physical degrees of freedom of the Maxwell field satisfy the same  equation as massless minimally coupled
scalar fields. However, the renormalized $\langle T_{\mu \nu}\rangle_{\text {ren}}^{\text {Maxwell}}$  is clearly different from $2\langle T_{\mu \nu}\rangle_{\text {ren}}^{\xi=1/6 \ \text {scalar}}$. Similar conclusions could be expected for spin-$2$ fields.

{\it Acknowledgements.}\ 
 This work is supported by the Spanish Grants No. PID2020-116567GBC2-1  and No. PROMETEO/2020/079 (Generalitat Valenciana).  F. J. Mara\~{n}\'on-Gonz\'alez is supported by the Ministerio de Ciencia, Innovaci\'on y Universidades, Ph.D. fellowship, Grant No. FPU22/02528. 
Some of the computations have been done with the help of \textit{Mathematica}\texttrademark.

\appendix{} \label{apA}

 Here we give the WKB expansion of the integrand of  (\ref{rhoomega}), (\ref{Pomega}), (\ref{rholambda}) and (\ref{Plambda}) 
 
 \begin{equation}
    \braket{\rho}_\Omega = \braket{T^0_0}_\Omega = \int \frac{d^3 k}{16 \pi^3 a^3}\left(\frac{\dot a \dot \Omega }{2 a
   \Omega ^2}+\frac{\dot a^2}{4
   a^2 \Omega
   }+\frac{\omega^2}{\Omega
   }+\frac{\dot \Omega
   ^2}{4 \Omega ^3}+\Omega  \right)
\end{equation}
\begin{equation}
    \braket{P}_\Omega = -\braket{T^1_1}_\Omega = \int \frac{d^3 k}{48 \pi^3 a^3}\left(\frac{\dot a \dot \Omega }{2 a
   \Omega ^2}+\frac{\dot a^2}{4
   a^2 \Omega
   }+\frac{\omega^2}{ \Omega
   }-\frac{2m^2}{\Omega }+\frac{\dot \Omega
   ^2}{4 \Omega ^3}+\Omega \right)
\end{equation}
\begin{equation}
\begin{split}
    \braket{\rho}_\Lambda = \braket{T^0_0}_\Lambda = \int \frac{d^3 k}{32 \pi^3 a^3}&\left(\frac{m^4 \dot a^2}{a^2 \Lambda
    \omega ^4}-\frac{m^2 \dot a
   \dot \Lambda }{a \Lambda ^2 \omega
   ^2}-\frac{3 m^2 \dot
   a^2}{a^2 \Lambda 
   \omega ^2}+\frac{3 \dot a \dot \Lambda
   }{2 a \Lambda ^2}+\frac{9
   \dot a^2}{4 a^2 \Lambda
   } \right. \\
   &\left.+\frac{\dot \Lambda ^2}{4 \Lambda
   ^3}+\frac{\omega ^2}{\Lambda }+\Lambda
     \right)
\end{split}
\end{equation}
\begin{equation}
\begin{split}
    \braket{P}_\Lambda = -\braket{T^1_1}_\Lambda = \int \frac{d^3 k}{96 \pi^3 a^3}&\left( -\frac{2 m^6 \dot a^2}{a^2 \Lambda
    \omega ^6}+\frac{2 m^4
   \dot a \dot \Lambda }{a \Lambda
   ^2 \omega ^4}+\frac{9 m^4
   \dot a^2}{a^2 \Lambda 
   \omega ^4}-\frac{6 m^2 \dot a \dot \Lambda
   }{a \Lambda ^2 \omega
   ^2}-\frac{27 m^2 \dot a^2}{2
   a^2 \Lambda  \omega ^2} \right. \\ 
   &\left. +\frac{9
   \dot a \dot \Lambda}{2 a
   \Lambda ^2}+\frac{27 \dot a^2}{4
   a^2 \Lambda }-\frac{m^2 \dot \Lambda
   ^2}{2 \Lambda ^3 \omega ^2}-\frac{2
   m^2 \Lambda }{\omega ^2}+\frac{3 \dot \Lambda
   ^2}{4 \Lambda ^3}-\frac{\omega
   ^2}{\Lambda }+3 \Lambda 
     \right) \ .
\end{split}
\end{equation}

\appendix{} \label{apB}
We provide the adiabatic subtractions for a $\xi$-generic scalar field, in cosmic time. See \cite{bunch80} for a more complete approach which includes the formal expressions, but in conformal time. Below, $\tilde \xi \equiv \xi - 1/6$; so that $\tilde \xi = 0$ for a conformal scalar, and $\tilde \xi = -1/6$ for a minimal scalar. Note that finite integrals have been evaluated explicitly. 
\begin{equation}
\begin{split}
    \braket{\rho}_\text{scalar}^{(0-4)} &= -\frac{\tilde \xi \ddot a^2}{16 \pi ^2
   a^2}+\frac{\ddot a^2}{960 \pi ^2
   a^2}+\frac{m^2 \dot a^2}{96 \pi ^2
   a^2}+\frac{9 \tilde \xi^2 \dot a^4}{4 \pi ^2
   a^4}-\frac{3 \tilde \xi \dot a^4}{16 \pi ^2
   a^4}+\frac{\dot a^4}{480 \pi ^2
   a^4}\\
   &+\frac{\tilde \xi \dddot a \dot a}{8 \pi
   ^2 a^2}-\frac{\dddot a  \dot a}{480 \pi ^2
   a^2}+\frac{9 \tilde \xi^2 \dot a^2 \ddot a}{4
   \pi ^2 a^3}+\frac{\tilde \xi \dot a^2 \ddot a}{8
   \pi ^2 a^3}-\frac{\dot a^2 \ddot a}{480 \pi ^2
   a^3} \\
   &+ \int \frac{d^3 k}{16 \pi^3 a^3} \left( \frac{9 \tilde \xi^2 \ddot a^2}{2 a^2 \omega
   ^3}-\frac{3 m^2 \tilde \xi \dot a^2}{a^2
   \omega ^3}+\frac{27 \tilde \xi^2 \dot a^4}{2 a^4
   \omega ^3}-\frac{3 \tilde \xi \dot a^2}{a^2 \omega
   }\right.\\
   &\left.-\frac{9 \tilde \xi^2 \dddot a
   \dot a}{a^2 \omega ^3}-\frac{9 \tilde \xi^2
   \dot a^2 \ddot a}{a^3 \omega ^3}+\omega  \right) \ , 
\end{split}
\end{equation}
\begin{equation}
\begin{split}
    \braket{P}_\text{scalar}^{(0-4)} &= -\frac{\tilde \xi \ddddot a}{24 \pi ^2
   a}+\frac{\ddddot a}{1440 \pi ^2
   a}-\frac{m^2 \ddot a}{144 \pi ^2
   a}-\frac{9 \tilde \xi^2 \ddot a^2}{8 \pi ^2
   a^2}-\frac{\tilde \xi \ddot a^2}{16 \pi ^2
   a^2}+\frac{\ddot a^2}{960 \pi ^2
   a^2}\\
   &-\frac{m^2 \tilde \xi \dot a^2}{4 \pi ^2
   a^2}-\frac{m^2 \dot a^2}{288 \pi ^2
   a^2}+\frac{15 \tilde \xi^2 \dot a^4}{8 \pi ^2
   a^4}-\frac{\tilde \xi \dot a^4}{16 \pi ^2
   a^4}+\frac{\dot a^4}{1440 \pi ^2 a^4}-\frac{3
   \tilde \xi^2 \dddot a \dot a}{2 \pi ^2
   a^2}\\
   &-\frac{\tilde \xi \dddot a \dot a}{12 \pi
   ^2 a^2}+\frac{\dddot a \dot a}{720 \pi ^2
   a^2}-\frac{15 \tilde \xi^2 \dot a^2 \ddot a}{4
   \pi ^2 a^3}+\frac{\tilde \xi \dot a^2 \ddot a}{4
   \pi ^2 a^3}-\frac{\dot a^2 \ddot a}{360 \pi ^2
   a^3}\\
   &+ \int \frac{d^3 k}{48 \pi^3 a^3} \left( \frac{9 \tilde \xi^2 \ddddot a}{a \omega ^3}+\frac{6
   m^2 \tilde \xi \ddot a}{a \omega
   ^3}+\frac{27 \tilde \xi^2 \ddot a^2}{2 a^2 \omega
   ^3}+\frac{6 \tilde \xi \ddot a}{a \omega
   }+\frac{27 \tilde \xi^2 \dot a^4}{2 a^4 \omega
   ^3}\right.\\
   &\left.-\frac{3 \tilde \xi \dot a^2}{a^2 \omega
   }+\frac{18 \tilde \xi^2 \dddot a
   \dot a}{a^2 \omega ^3}-\frac{54 \tilde \xi^2
   \dot a^2 \ddot a}{a^3 \omega
   ^3}-\frac{m^2}{\omega }+\omega  \right) \ . 
\end{split}
\end{equation}


\begin{thebibliography}{9}

\bibitem{Parker} L. Parker, Phys. Rev. Lett. {\bf 21}, 562 (1968); Phys. Rev. {\bf 183}, 1057 (1969); Phys. Rev. D {\bf 3}, 346 (1971).  

\bibitem{kolb-long}
E.~W.~Kolb and A.~J.~Long, {\it Cosmological gravitational particle production and its implications for cosmological relics}, 
arXiv:2312.09042.

\bibitem{ford}
L.~H.~Ford,
Rept. Prog. Phys. \textbf{84}, 116901 (2021).

\bibitem{parker-fulling} L.~Parker and S.~A.~Fulling,  Phys.~Rev.~D {\bf 9}, 341 (1974). 

\bibitem{parker-fulling2} S. A. Fulling and L. Parker, Ann.~Phys. (N.Y.) {\bf 87}, 176 (1974); S. A. Fulling, L. Parker, and B. L. Hu, Phys. Rev. D {\bf 10}, 3905 (1974).

\bibitem{parker12} L. Parker,  J.~Phys.~A {\bf 45}, 374023 (2012).

\bibitem{birrell-davies} N.~D.~Birrell  and P.~C.~W.~Davies, {\it Quantum Fields in Curved Space} (Cambridge University Press, Cambridge, England 1982).

\bibitem{fulling} S.~Fulling, {\it Aspects of Quantum Field Theory in Curved Space-Time} (Cambridge University Press, Cambridge, England 1989).

\bibitem{parker-toms}L.~Parker and D.~J.~Toms, {\it Quantum Field Theory in Curved Spacetime: Quantized Fields
and Gravity}, Cambridge University Press, Cambridge, England (2009).

\bibitem{hu-verdaguer} B.-L. B. Hu and E. Verdaguer, {\it Semiclassical and Stochastic Gravity: Quantum Field Effects on Curved Spacetime} (Cambridge  University Press, Cambridge, England 2020).


\bibitem{ZS} Y.B. Zeldovich and A.A. Starobinsky, Sov. Phys. JETP {\bf 34}, 1159 (1972).




\bibitem{BPHZ} W. Zimmerman, {\it Local operator product and renormalization in quantum field theory}, pp 395-589 of Lectures on Elementary Particles and Quantum Field Theory. Vol. 1. /Deser, Stanley (ed.). Cambridge, Mass. Massachusetts Inst. of Tech. Press (1970).

\bibitem{BPHZ2} K. Sibold (2010), {\it Bogoliubov-Parasiuk-Hepp-Zimmermann renormalization scheme}, Scholarpedia 5(5):7306 (2010). 	

\bibitem{Hu} B. L. Hu, Phys. Lett. A {\bf 71}, 169 (1979); Phys. Rev. D {\bf 18}, 4460 (1978).

\bibitem{Birrell78} N. D. Birrell,  Proc. R. Soc. Lond. B. {\bf 361}, 513 (1978).


\bibitem{bunch80} T. S. Bunch, J. Phys. A {\bf 13}, 1297 (1980). 

\bibitem{Anderson-Parker} P. R. Anderson and L. Parker, Phys. Rev. D {\bf 36}, 2963 (1987).

\bibitem{Anderson} P.~R.~Anderson and W.~Eaker, Phys. Rev. D {\bf 61}, 024003 (1999)

\bibitem{Habit} S.~Habib, C.~Molina-Paris, and E.~Mottola,  Phys.~Rev.~D {\bf 61}, 024010 (1999)





\bibitem{agullo-navarro-olmo-parker}
I.~Agullo, J.~Navarro-Salas, G.~J.~Olmo and L.~Parker,
Phys. Rev. D \textbf{84}, 107304 (2011)


\bibitem{Bavarsad:2016cxh}
E.~Bavarsad, C.~Stahl and S.~S.~Xue,
Phys. Rev. D \textbf{94}, 104011 (2016)



\bibitem{Bavarsad:2017oyv}
E.~Bavarsad, S.~P.~Kim, C.~Stahl and S.~S.~Xue,
Phys. Rev. D \textbf{97}, 025017 (2018)



\bibitem{Moreno-Pulido20}
C.~Moreno-Pulido and J.~Sola,
Eur. Phys. J. C \textbf{80}, 692 (2020).


\bibitem{Sola22} J.~Sola Peracaula,
Phil. Trans. Roy. Soc. Lond. A \textbf{380}, 20210182 (2022).

\bibitem{Silvia22} S.~Pla and E.~Winstanley,
Phys. Rev. D \textbf{107}, 025004 (2023).



\bibitem{ferreiro-torrenti} 
A.~Ferreiro, S.~Monin and F.~Torrenti,
Phys. Rev. D \textbf{109},  045015 (2024).

\bibitem{silvia}
S.~Pla and B.~A.~Stefanek,
Phys. Lett. B \textbf{856}, 138926 (2024).


\bibitem{Hu23}
Y.~C.~Xie, S.~Butera and B.~L.~Hu,
Comptes Rendus Physique \textbf{25}, no.S2, 1-22 (2024).


\bibitem{LNT} A.~Landete, J.~Navarro-Salas, and F.~Torrenti, Phys~Rev.~D {\bf 88}, 061501 (2013); {\bf 89}, 044030 (2014).

\bibitem{RNT} A.~del Rio, J.~Navarro-Salas, and F.~Torrenti,  Phys.~Rev.~D {\bf 90}, 084017 (2014).

\bibitem{Ghosh15}
S.~Ghosh,
Phys. Rev. D \textbf{91},  124075 (2015); {\bf 93},   044032 (2016).


\bibitem{Yukawa} A. del Rio, A. Ferreiro, J. Navarro-Salas, F. Torrenti, 
{\it Phys. Rev. D} {\bf 95}, 105003 (2017).




\bibitem{BFNV} J. F. Barbero G., A. Ferreiro, J. Navarro-Salas, and E. J. S. Villase\~nor, {\it Phys. Rev. D} {\bf  98}, 025016 (2018).




\bibitem{Ferreiro18}
A.~Ferreiro and J.~Navarro-Salas,
Phys. Rev. D \textbf{97}, no.12, 125012 (2018).


\bibitem{Ferreiro18b}
A.~Ferreiro, J.~Navarro-Salas and S.~Pla,
Phys. Rev. D \textbf{98}, no.4, 045015 (2018).






\bibitem{BNP} P. Beltran-Palau, J. Navarro-Salas, and S. Pla, {\it Phys. Rev. D} {\bf 101}, 105014 (2020).


\bibitem{Silvia-Paul} S.~Pla, I.~M.~Newsome, R.~S.~Link, P.~R.~Anderson and J.~Navarro-Salas,
Phys. Rev. D \textbf{103}, 105003 (2021).

\bibitem{Eduardo-Luis} \'A.~\'Alvarez-Dom\'\i{}nguez, L.~J.~Garay, E.~Mart\'\i{}n-Mart\'\i{}nez and J.~Polo-G\'omez,
Phys. Rev. Lett. \textbf{133},  041401 (2024)

\bibitem{L76} J. Lowenstein, Commun. Math. Phys. {\bf 47}, 53 (1976).

\bibitem{Ferreiro-Navarroplb} A.~Ferreiro and J.~Navarro-Salas,
Phys. Lett. B \textbf{792}, 81-85 (2019).

\bibitem{chimento-cossarini} L.P. Chimento and A. E. Cossarini, {\it Phys. Rev. D} {\bf 41}, 3101 (1990).



\bibitem{agullo} I. Agullo, A. Landete, and J. Navarro-Salas, Phys. Rev. D {\bf 90}, 124067 (2014).

\bibitem{chu-koyama} C.-S. Chu and Y. Koyama,  Phys. Rev. D {\bf 95}, 065025 (2017).

\bibitem{ordines-carlson} T. M. Ordines and  E.D. Carlson, Phys. Rev. D {\bf 106}, 103537 (2022).

\bibitem{Javier-Pepe23}
F.~J.~Mara\~n\'on-Gonz\'alez and J.~Navarro-Salas,
Phys. Rev. D \textbf{108}, 125001 (2023).




\bibitem{Peskin}M.~E.~Peskin and D.~V.~Schroeder,
{\it An Introduction to Quantum Field Theory},
(Addison-Wesley, 1995).


\bibitem{piguet-sorella} O. Piguet and S.P. Sorella, {\it Algebraic renormalization: perturbative renormalization, symmetries and anomalies}, Lec. Notes Phys. {\bf M28} (Springer-Verlag, 1995).

\bibitem{pirk} K.-T. Pirk,  Phys. Rev. D \text{bf 48}, 3779 (1993).

\bibitem{sergi} S.~Nadal-Gisbert, J.~Navarro-Salas and S.~Pla,
Phys. Rev. D \textbf{107},  085018 (2023).

\bibitem{parker-wang} L.~Parker and Y.~Wang,
Phys. Rev. D \textbf{39}, 3596-3605 (1989).

\bibitem{Duff} M.~J.~Duff,
Class. Quant. Grav. \textbf{11}, 1387 (1994).

\bibitem{buchbinder-shapiro21} I.L. Buchbinder and I.L. Shapiro, {\it Introduction to Quantum Field Theory with Applications to Quantum Gravity} (Oxford University Press, Oxford, England 2021).

\bibitem{buchbinder-berredo-shapiro} I.L. Buchbinder G. de Berredo-Peixoto, and I.L. Shapiro,     {\it Phys. Lett. B} {\bf 649}, 454 (2007).

\bibitem{allen-folacci} B. Allen and A. Folacci, {\it Phys. Rev. D} {\bf 35} 3771 (1987).




\end{thebibliography}
\end{document}